\documentstyle[aps,12pt]{revtex}

\begin{document}

\preprint{EFUAZ FT-97-54}

\title{A Note on the Neutrino Theory of Light\thanks{Submitted to
``Speculations in Science and Technology"}}

\author{{\bf Valeri V. Dvoeglazov}}

\address{Escuela de F\'{\i}sica, Universidad Aut\'onoma de Zacatecas\\
Apartado Postal C-580, Zacatecas 98068 Zac., M\'exico\\
Internet address: valeri@cantera.reduaz.mx\\
URL: http://cantera.reduaz.mx/\~~valeri/valeri.htm
}

\date{November 24, 1997}

\maketitle

\medskip

\begin{abstract}
In this small note we ask several questions which
are relevant to the construction of the self-consistent
neutrino theory of light. The previous confusions in such attempts
are explained in the more detailed publication.
\end{abstract}

\medskip

The basic idea of the de Broglie-Jordan neutrino theory of
light~\cite{DeBroglie}-\cite{Campo} is: on using the $(1/2,0)\oplus
(0,1/2)$ field operators to construct the field operator of the
$(1/2,1/2)$ field, the 4-vector potential, which is used for describing
the light. Due to the Pryce theorem~\cite{Pryce} we know that it is
impossible to use the solutions of the Dirac equation in order to obtain
the transverse components of the 4-vector potential. As shown by Barbour
{\it et al.} in the most elegant way, ref.~\cite{Barbour}, the Jordan's
{\it anzatz}
\begin{equation}\label{ansatz}
{\cal  A}_{\mu} (k)
= \int_0^1 f (\lambda) \overline \psi ((\lambda -1) k) \gamma_{\mu} \psi
(\lambda k) d\lambda \,\,
\end{equation}
reduces to
\begin{equation}
{\cal A}_\mu (k)
= -ik_\mu \int_0^1 f(\lambda) \varphi (k, \lambda) d\lambda\,\, ,
\end{equation}
with
\begin{equation}
\varphi (k,\lambda) = \lim_{m\rightarrow 0}
[{1\over m} \overline\psi ((\lambda -1) k) \psi (\lambda k) ]\,\,
\end{equation}
being well-defined in the massless limit.
In the above formulas we used the notation: ${\cal A}_\mu (k)$ is the
electromagnetic potential in the momentum space; $k$ is the photon
four-momentum; $\gamma_\mu$ is the usual $4\times 4$ matrices and $\psi$ is
the spin-1/2 states;  $f (\lambda)$ is some suitable weight
function.  In the de Broglie theory one has $f (\lambda) =
\delta (\lambda -a)$, $a$ is some constant.

In this connection I have several questions:

\begin{itemize}

\item
Why do we use the charged particle
(electron-positron) 4-spinors in order to construct the photon states?
Indeed, the $\psi (\lambda k)$ was assumed to be solutions of the
{\it Dirac equation}. It is natural to call the previous attempts
as ``the electron theory of light"! But, if we want to construct
the neutrino theory of light one should use the field operator composed
of self/anti-self charge conjugate states, the states which are the
eigenstates of the charge conjugation operator $S^c_{[1/2]} = {\cal
C}_{[1/2]} {\cal K}$, and not of the charge operator $Q$. Various
forms of the operators composed from such states have been proposed
recently by Ahluwalia~\cite{DVA} and myself~\cite{DVO1,DVO2,DVO3,DVO4}.
For instance the equation (46) of~\cite{DVA}, the massive neutrino
field operator,\footnote{The present-day experimental data indicate
the existence of the mass of neutrino, e.~g., ref.~\cite{exp}.}
reads
\begin{eqnarray} \nu ^{^{DL}} (x) &\equiv& \int {d^3 {\bf p} \over
(2\pi)^3} \,\, {1\over 2p_0} \times\nonumber\\ &&\sum_\eta [\lambda^S_\eta
(p^\mu) a_\eta (p^\mu) \exp (-ip\cdot x) +\lambda^A_\eta (p^\mu)
b_\eta^\dagger (p^\mu) \exp (+ip\cdot x) ]\,\, .
\end{eqnarray}
These
states are the solutions of different equations; e.~g., for the
positive-energy solutions one has
\begin{mathletters} \begin{eqnarray}
i \gamma^\mu \partial_\mu \lambda^S (x) - m \rho^A (x) &=& 0 \quad,
\label{11}\\ i \gamma^\mu \partial_\mu \rho^A (x) - m \lambda^S (x) &=& 0
\quad, \label{12}
\end{eqnarray} \end{mathletters}
where $\lambda^{S,A}$
and $\rho^{A,S}$ are the corresponding Fourier transforms of the
self/anti-self charge conjugate 4-spinors constructed on the basis of
$\phi_{_{L,R}} (p^\mu) $ 2-spinors, respectively.  The 4-spinors of the
second kind differ in the parity properties from the Dirac 4-spinors. The
origin of the possibility to have various forms of the field operators
(answering to different physical content) in the same representation of
the Lorentz group was noted and clarified in the forgotten lectures of
Wigner~\cite{Wig}.

\item
I do not understand why physicists were so unpleasant to accept that
the 4-vector potential might be longitudinal (in the sense $h=0$). In any
theory based on the Lorentz group representation $D(A,B)$ the relation
$B-A = h$ must be satisfied~\cite{Wein}.  In this formula $h$ refers to
the helicity of physical states.  If one uses the representation
$D(1/2,1/2)$, the physical field must have the helicity $h=0$. In fact, we
came across with this theory for the 4-vector potential when we tried to
understand the numerous confusions in the theory of the so-called
Kalb-Ramond field, see our recent paper~\cite{DVO5}. Our consideration
predicts that the transversality/longitudinality of the 4-vector potential
in the $(1/2,1/2)$ representation depends on the chosen normalization. The
explicit forms of the 4-vectors in the momentum representation are
\begin{mathletters}
\begin{eqnarray}
u^\mu ({\bf p}, +1)= -{N\over \sqrt{2}m}\pmatrix{p_r\cr m+ {p_1 p_r \over
E_p+m}\cr im +{p_2 p_r \over E_p+m}\cr {p_3 p_r \over
E_p+m}\cr}&\quad&,\quad u^\mu ({\bf p}, -1)= {N\over
\sqrt{2}m}\pmatrix{p_l\cr m+ {p_1 p_l \over E_p+m}\cr -im +{p_2 p_l \over
E_p+m}\cr {p_3 p_l \over E_p+m}\cr}\quad,\quad\\ u^\mu ({\bf p}, 0) &=&
{N\over m}\pmatrix{p_3\cr {p_1 p_3 \over E_p+m}\cr {p_2 p_3 \over
E_p+m}\cr m + {p_3^2 \over E_p+m}\cr}\quad, \quad
\end{eqnarray}
\end{mathletters}
($N=m$ is the normalization and $p_{r,l} = p_1 \pm ip_2$).
They do not
diverge in the massless limit and can be used to obtain the
physical strengths ${\bf E}$ and ${\bf B}$ (which are also have
good behavior in the massless limit).  Two of
the massless functions (with $h = \pm 1$) are equal to zero when the
particle, described by this field, is moving along the third axis ($p_1 =
p_2 =0$,\, $p_3 \neq 0$).  The third one ($h = 0$) is \begin{equation}
 u^\mu (p_3, 0) \mid_{m\rightarrow 0} = \pmatrix{p_3\cr 0\cr 0\cr {p_3^2
 \over E_p}\cr} \equiv  \pmatrix{E_p \cr 0 \cr 0 \cr E_p\cr}\quad,
\end{equation}
and at the rest ($E_p=p_3 \rightarrow 0$) also vanishes. For the first
time, the matters related with the correct choice of the physical
normalization have been considered in ref.~\cite{DVA0}.

Thus, the field operator composed of (6,7)
describes the `longitudinal photons',
what is in the complete accordance with the Weinberg theorem $B-A= h$.
For the sake of completeness let us present the potential and fields
corresponding to the ``time-like" polarization:
\begin{equation} u^\mu
({\bf p}, 0_t) = {N \over m} \pmatrix{E_p\cr p_1 \cr p_2\cr
p_3\cr}\quad,\quad {\bf B}^{(\pm)} ({\bf p}, 0_t) = {\bf 0}\quad,\quad
{\bf E}^{(\pm)} ({\bf p}, 0_t) = {\bf 0}\quad.\quad
\end{equation}
The polarization vector $u^\mu ({\bf p}, 0_t)$ has the good behavior in
$m\rightarrow 0$, $N=m$ (and also in the subsequent limit ${\bf p}
\rightarrow {\bf 0}$) and it may also correspond to some quantized field
(particle).  Furthermore, in the case of the normalization of potentials
to the mass $N=m$  the physical fields which correspond to the
``time-like" polarization  are equal to zero identically.  The
longitudinal fields (strengths) are equal to zero in this limit only when
one chooses the frame  with $p_3 = \mid {\bf p} \mid$, cf. with the light
front formulation, ref.~\cite{DVALF}.  In the case $N=1$ we have, in
general, the divergent behavior of potentials and strengths. Usually, the
divergent part of the potentials was referred to the gauge-dependent part.
But, according to the Weinberg theorem it is the gauge part which is the
{\it physical} field.

\item
Why do the physicists not want to use the electric and magnetic field
${\bf E}$ and ${\bf B}$ in order to obtain the transverse components of
the light?\footnote{Unfortunately,  the papers~\cite{Sen,Campo}
did not practically draw any attention.} In fact one can propose
the analogues of the Jordan's {\it anzatz} for the antisymmetric tensor
field of the second rank:
\begin{mathletters}
\begin{eqnarray} F_{\mu\nu}
&=& \int_0^1 f (\lambda) \overline \psi_{\bar\nu} ((\lambda -1) k)
\sigma_{\mu\nu} \psi_{\nu} (\lambda k)d \lambda \,\,,\\
\widetilde
F_{\mu\nu} &=&  \int_0^1 f (\lambda) \overline \psi_{\bar\nu}  ((\lambda
-1) k) \widetilde \sigma_{\mu\nu} \psi_{\nu} (\lambda k) d\lambda \,\,,
\end{eqnarray}
\end{mathletters}
with $\sigma_{\mu\nu} = {i\over 2} [\gamma^\mu ,
\gamma^\nu ]_-$ and $\widetilde\sigma_{\mu\nu}$ being its dual.

\end{itemize}

I think that on the basis of this three observations one can explain many
interpretational confusions in the neutrino theory of light and obtain
both transverse and longitudinal components of the 4-vector potential
from various types of the field operators. The detailed explanation
was presented in another publication submitted to ``Nuovo Cimento A".

{\it Acknowledgments.} I acknowledge the help of Prof. A. F. Pashkov,
who, in fact, began the work in this direction.

Zacatecas University, M\'exico, is thanked for awarding
the full professorship.  This work has been partly supported by
the Mexican Sistema Nacional de Investigadores, the Programa de Apoyo a la
Carrera Docente and by the CONACyT, M\'exico under the research project
0270P-E.


\begin{references}

\footnotesize{
\baselineskip12pt

\bibitem{DeBroglie} L. de Broglie, {\it C. R.  Acad. Sci. (Paris)}
{\bf 195} (1932) 536, 577, 862; {\it ibid}
{\bf 197} (1933) 1337; {\it ibid} {\bf 198} (1934) 135;
{\it ibid} {\bf 199} (1934) 445, 1165; {\it ibid} {\bf 230} (1950)
1434; {\it J. Phys. Rad.} {\bf 12} (1951) 509; L. de Broglie and J.
Winter, {\it C. R. Acad. Sci. (Paris)} {\bf 199} (1934) 813; L. de Broglie
and M. A. Tonella, {\it ibid} {\bf 230} (1950) 1329.

\bibitem{Jordan} P. Jordan,  {\it Egreb. Exakt. Naturw.}
{\bf 7}  (1928) 158; {Z. Phys.}  {\bf 93} (1935) 464;
{\it ibid} {\bf 98} (1936) 759; {\it ibid} {\bf 99} (1936) 109;
{\it ibid} {\bf 102} (1936) 243; {\it ibid} {\bf 105} (1937) 114, 229;
P. Jordan and R. de L. Kronig, {\it ibid} {\bf 100}
(1936) 569.

\bibitem{Scherzer} O. Scherzer, {\it Z. Phys.} {\bf 97} (1935) 725.

\bibitem{Kronig} R. de L. Kronig, {\it Physica} {\bf 2} (1935) 491, 854, 968;
{\it ibid} {\bf 3} (1936) 1120; {\it Nature} {\bf 137} (1936) 149.

\bibitem{Nath} M. Born and N. S. Nagendra Nath, {\it Proc. Indian Acad. Sci.}
{\bf 3} (1936) 318; {\it ibid} {\bf 4} (1936) 611;
N. S. Nagendra Nath,  {\it ibid} {\bf 3} (1936) 448.

\bibitem{Fock} V. A. Fock, {\it Phys. Z. Sowjet.} {\bf 11} (1937) 1;
{\it C. R. Acad. Sci. U. R. S. S.} {\bf 4} (1937) 229.

\bibitem{Sokolov} A. A. Sokolow, {\it Phys. Z. Sowjet.}
{\bf 12} (1937) 148.

\bibitem{Pryce} M. H. L. Pryce, {\it Proc. Roy. Soc. A}{\bf 165}
(1938) 247.

\bibitem{Case} K. M. Case, {\it Phys. Rev.} {\bf 106} (1957) 1316.

\bibitem{Barbour} I. M. Barbour, A. Bietti and B. F. Touschek,
{\it Nuovo Cim.} {\bf 28} (1963) 452.

\bibitem{Sen} D. K. Sen, {\it Nuovo Cim.} {\bf 31} (1964) 660.

\bibitem{Ferretti} B.  Ferretti, {\it Nuovo Cim.} {\bf 33} (1964) 264;
B. Ferretti and I. Venturi, {\it Nuovo Cim.} {\bf 35} (1964) 644.

\bibitem{Perkins} W. A. Perkins, {\it Phys. Rev.} {\bf 137} (1965) B1291;
{\it ibid} {\bf D5} (1972) 1375.

\bibitem{Bandy} P. Bandyopadhyay, {\it Nuovo Cim.} {\bf 38}
(1965) 1912.

\bibitem{Berezinsky} V. S. Berezinski\u{\i}, {\it ZhETF} {\bf 51}
(1966) 1374 [English translation: {\it Sov. Phys. JETP} {\bf 24}
(1965) 927].

\bibitem{Broido} M. M. Broido, {\it Phys. Rev.} {\bf 157} (1967) 1444.

\bibitem{Bandy2} P. Bandyopadhyay, {\it Phys. Rev.} {\bf 173} (1968) 1481;
{\it Nuovo Cim.} {\bf 55A} (1968)  367.

\bibitem{Inoue} H. Inoue, T. Tajima and S. Tanaka, {\it Progr. Theor. Phys.}
{\bf 48} (1972) 1338.

\bibitem{Bandy3} P. Bandyopadhyay and
P. R. Chaudhuri, {\it Phys. Rev. D}{\bf 3} (1971) 1378.

\bibitem{Sarkar}
H. Sarkar, B. Bhattacharya and P. Bandyopadhyay, {\it Phys. Rev. D}{\bf
11} (1975)  935.

\bibitem{Strazhev} V. I. Strazhev, {\it Int. J. Theor.
Phys.} {\bf 16}  (1977) 111.

\bibitem{Luther} A. Luther and K. D.
Schotte, {\it Nucl. Phys. B}{\bf  242} (1984) 407.

\bibitem{Mickels} J. Mickelsson, {\it J. Math. Phys.} {\bf 26}
(1985) 2346.

\bibitem{Campo} A. A. Campolattaro, {\it Int. J. Theor. Phys.} {\bf 19}
(1980) 99, 127; {\it ibid} {\bf 29} (1990) 141, 477.

\bibitem{DVA} D. V. Ahluwalia, {\it Int. J. Mod. Phys.} A{\bf 11} (1996)
1855.

\bibitem{DVO1} V. V. Dvoeglazov,  {\it Rev.  Mex.  Fis.  (Suppl.  1 --
Proc.  XVIII Oaxtepec Symp.  on Nucl.  Phys.,  Jan.  1995)} {\bf 41}
(1994) 159; {\it Int. J. Theor. Phys.} {\bf 34} (1995) 2467.

\bibitem{DVO2} V. V. Dvoeglazov, {\it Nuovo Cim.} {\bf 108}A (1995) 1467.

\bibitem{DVO3} V. V. Dvoeglazov, {\it Different Quantum Field Constructs
in the $(1/2,0)\oplus (0,1/2)$ Representation.} Preprint EFUAZ FT-96-30
(hep-th/9609142),  Zacatecas, July 1996, accepted in {\it Mod. Phys. Lett.
A}

\bibitem{DVO4} V. V. Dvoeglazov, {\it Hadronic Journal} {\bf 20} (1997)
435.

\bibitem{exp} A. Athanassopoulos {\it et al.}, {\it Phys. Rev. Lett.} {\bf
75} (1995) 2650.

\bibitem{Wig} E. P. Wigner, in {\it Group theoretical concepts
and methods in elementary particle physics -- Lectures of the Istanbul
Summer School of Theoretical Physics, 1962.} Ed. F. G\"ursey, (Gordon \&
Breach, 1965), p. 37.

\bibitem{Wein} S. Weinberg, {\it Phys. Rev. B}{\bf 134} (1964) 882.

\bibitem{DVO5} V. V. Dvoeglazov, {\it On the Importance of the
Normalization.} Preprint EFUAZ FT-96-39-REV, Zacatecas, Nov. 1997.

\bibitem{DVA0} D. V. Ahluwalia {\it et al.}, {\it Phys. Lett. B}{\bf 316}
(1993) 102. This paper presents an explicit example of the theory of the
Bargmann-Wightman-Wigner type in the $(1,0)\oplus (0,1)$ representation.

\bibitem{DVALF} D. V. Ahluwalia and M. Sawicki, Phys. Rev. D{\bf 47}
(1993) 5161.

}

\end{references}
\end{document}